\pdfoutput=1
\documentclass[aps,prl,twocolumn,showpacs,floatfix]{revtex4-1}
\usepackage[utf8]{inputenc}
\usepackage{amsmath}
\usepackage{amsfonts}
\usepackage{graphicx}
\usepackage{braket}
\usepackage{xcolor}

\begin{document}
%\draft
%\preprint{{\bf ETH-TH/98-??}}

\title {Ising anyonic topological phase of interacting Fermions in one dimension}

\author{K. Guther}
\email{guther@itp3.uni-stuttgart.de}
\altaffiliation[Current address: ]{Max-Planck-Institute for Solid State
  Research, Heisenbergstra{\ss}e 1, 70569 Stuttgart, Germany}

\author{N. Lang}
\author{H. P. B\"uchler}

\affiliation{Institute for Theoretical Physics III and Center for Integrated Quantum Science and Technology, University of Stuttgart, Pfaffenwaldring 57, 70550 Stuttgart, Germany}

\date{\today}

\begin{abstract}
    We study a microscopic model of interacting fermions in a ladder setup, where the total number of particles is conserved.
    At a special point, the ground state is known %exactly 
    and gives rise to a topological state of matter 
    with edge modes obeying the  %non-abelian 
    statistics of Ising anyons.
    Using a combination of bosonization 
    as well as  full scale numerical density-matrix renormalization group analysis, 
    we map out the full phase diagram. 
    We find that the topological phase survives in an extended parameter regime. 
    Remarkably, an additional symmetry is required to protect the topological phase. 
\end{abstract}


\maketitle

The potential of low-dimensional quantum systems to exhibit topological order and host quasiparticles with anyonic statistics 
marks one of the most fascinating phenomena in condensed matter physics. The theoretical description of such systems \cite{thouless, wen} 
and their potential applications to fault-tolerant quantum computation \cite{kitaev_computation, nayak} 
sparked the search for their realization in condensed matter setups and cold atomic gases. 
Prominent examples featuring non-abelian anyons are topological $p$-wave superconductors 
within their mean-field description \cite{ivanov, kitaev_wires, asashi},  for which the existence of Majorana bound states in one \cite{kitaev_wires} 
and two dimensions \cite{green, ran} has been shown. Notably, recent experiments revealed signatures consistent with Majorana zero-energy 
edge-modes in nanowires \cite{rokhinson, das, deng, albrecht, mourik, nadj-perge}. However, a complete picture of Majorana modes in a particle conserving setup (i.e., beyond mean-field),
and its competition with the gapless Goldstone mode in one-dimension, remains an open challenge. Especially for cold atomic gases, 
the need for a microscopic, number-conserving theory is expected to be of paramount importance.

Several theoretical studies have focused on the understanding of Majorana-like edge modes 
in one-dimensional systems beyond mean field theory. A prominent approach is based on effective low-energy 
theories employing bosonization \cite{cheng,lutchyn,halperin,ruhman,keselman},
while the first exact solution has been restricted to models requiring unphysical long-range interactions \cite{ortiz}. 
Alternatively, one-dimensional double wire setups of spinless fermions with pair-correlated hopping 
have emerged as promising candidates: First, numerical studies have demonstrated the appearance of edge states 
in such a model \cite{kraus}; later, exactly solvable extensions have been found \cite{lang,iemini}
and the Ising anyonic statistics of the edge states was demonstrated microscopically \cite{lang}. At this exactly solvable point,
the topological properties are protected in generic wire networks by particle number conservation alone. However, it is not yet established 
whether this topological state extends over a finite parameter regime, and whether particle conservation is enough to protect the 
topological phase in general; the latter is of special interest as previous approaches by bosonization revealed an algebraic splitting \cite{lutchyn,ruhman}
in absence of a local symmetry.

%% ++++++++++++++++++++++++++++++++++++++++++++++++++++++++++++++++++++++++++++++++++++++++++++++++++++++++++++++++
\begin{figure}[t]
    \centering
    \includegraphics[width=0.95\columnwidth]{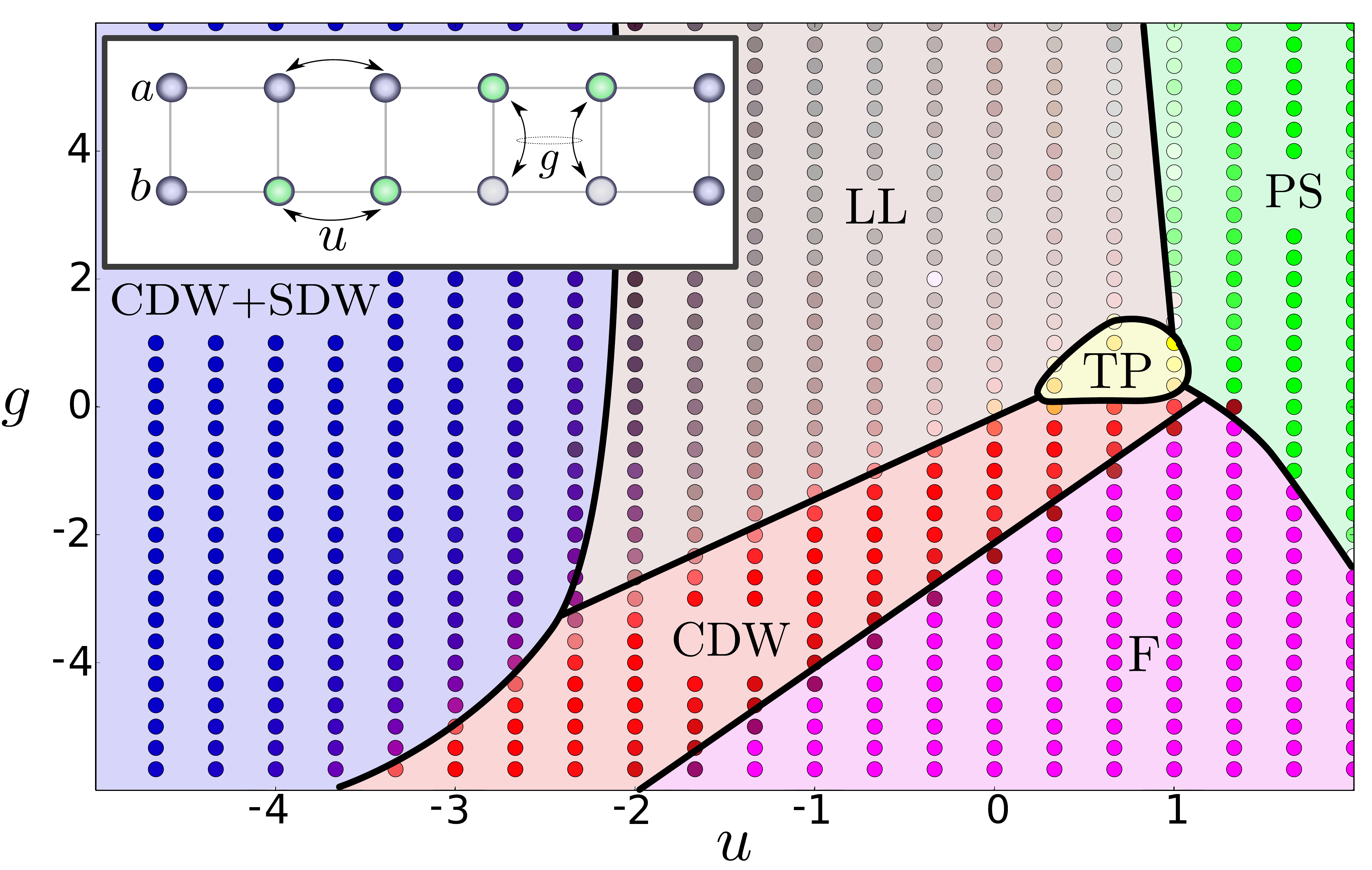}
    \caption{%
        \textit{Phase diagram.}
        Numerical results for the phase diagram at half filling with system length $L=50$ and $w=1$. 
        The  dots represent pairs $(u,g)$ for which the ground state was computed.
        The color encodes the respective phase:
        %The parameters are given in Eq.~(\ref{Hamiltonian});
        %dots represent pairs $(u,g)$ for which the ground state was computed numerically.
        %The color encodes the respective phase:
        (PS) Intra-chain phase separation,
        (F) Inter-chain phase separation,
        (LL) Gapless Luttinger liquid with algebraic correlations,
        (TP) Topological phase with $p$-wave pairing,
        (CDW) Charge density wave due to a gap in the symmetric sector,
        (CDW+SDW) Charge- and spin density waves due to gaps in both sectors.
        The inset illustrates the double wire setup, highlighting the intra-
        and inter-chain density-density interactions. The dotted line
        indicates the extent of the topological phase at filling
        $\rho=0.35$. Then it includes the point $g=u=0$ studied in
        Ref.~\cite{kraus}. For a quantitative descritpion of the color code,
        see the appendix.
    }
    \label{fig:PD}
\end{figure}
%% ++++++++++++++++++++++++++++++++++++++++++++++++++++++++++++++++++++++++++++++++++++++++++++++++++++++++++++++++

In this work, we derive the full ground state phase diagram of a microscopic model with spinless fermions on a double wire setup
which contains both the analytically solvable Hamiltonian from \cite{lang} and the Hamiltonian employed in \cite{kraus} as special cases. 
Using the combination of bosonization for a qualitative description of the phase diagram 
as well as a full scale numerical density-matrix renormalization group (DMRG) analysis,  
we bridge the gap between the previous approaches, present the full phase diagram
and demonstrate the stability of the topological phase.  In particular, we find that the previously studied 
models belong to the same phase and are smoothly connected to each other.  
We  investigate the stability of the topological phase under perturbations. Especially, we find that  perturbations
violating the subchain parity and time-reversal symmetry lead to an algebraic splitting of the edge modes. 
This implies that even in a particle conserving setting the topological phase still belongs to the class 
which can be protected by either subchain parity or time-reversal symmetry.

%% ++++++++++++++++++++++++++++++++++++++++++++++++++++++++++++++++++++++++++++++++++++++++++++++++++++++++++++++++
\paragraph{Model:} 
%% ++++++++++++++++++++++++++++++++++++++++++++++++++++++++++++++++++++++++++++++++++++++++++++++++++++++++++++++++

We start with the microscopic Hamiltonian describing spinless fermions on a one-dimensional double chain. 
The chain consist of $L$ lattice sites, and  $a^\dagger_i$ ($b^\dagger_i$) denotes the fermionic 
creation operators on the upper (lower) chain at lattice site $i$. 
The Hamiltonian involves nearest-neighbor hopping, interactions within each chain  and interactions between the chains. 
It is conveniently written as 
\begin{equation}
    H= \sum_{i} 
    \sum_{\sigma\in \{a,b\}}\left[- A_{i}^{\sigma} + u\left( A_{i}^\sigma\right)^2 \right] + 
    \Big[ w  B_{i} + g B_{i}^2\Big] 
    \label{Hamiltonian}
\end{equation}
with the single-particle hopping within each chain
\begin{equation}
    A^{a} = a_i^\dagger a^{}_{i+1} + a^\dagger_{i+1}a^{}_i
    \quad\text{and}\quad
    A^{b} = b_i^\dagger b^{}_{i+1} + b^\dagger_{i+1}b^{}_i\,.
\end{equation}
The second term describes a nearest-neighbor attraction within each chain as well as a shift in 
chemical potential, i.e.,
\begin{displaymath}
    (A^{\sigma}_{i})^2= n^\sigma_{i} + n^\sigma_{i+1} - 2 n^\sigma_{i}n^\sigma_{i+1},
\end{displaymath}
with $\sigma\in\{a,b\}$. 
Here, $n^{\sigma}_i$ is the number of fermions on site $i$ of chain $\sigma$.
Furthermore, the interaction \textit{between} the chains is given by the pair-correlated hopping
\begin{equation}
    B_{i}=  a^\dagger_i a^\dagger_{i+1} b^{}_i b^{}_{i+1} + b^\dagger_i b^\dagger_{i+1} a^{}_i a^{}_{i+1}.
\end{equation}
Note that the sign of $w$ can be changed by the gauge transformation $a_j
\mapsto ia_j$. In the following, we hence restrict the analysis to $w>0$.
The last term in (\ref{Hamiltonian}) describes an inter-chain interaction and involves up to 
four-body interactions between the fermions
\begin{displaymath}
    B_{i}^2= 
    n^a_{i}n^a_{i+1} (1-n^{b}_{i}) (1-n^{b}_{i+1}) 
    +[\,a\leftrightarrow b\,]
\end{displaymath}
For simplicity, we fixed the particle tunneling within the chains to unity and
express all energies in terms of this tunneling energy. 
%
%The  interchain 
%pair-hopping of strength $W$ tends towards the formation of a $p$-wave pairing on each chain. 
%
In addition to the conservation of the total particle number $N$,
this Hamiltonian exhibits time-reversal symmetry (represented by complex conjugation) 
and conserves the fermionic subchain parity; the latter is denoted as $\alpha = \pm 1$ 
for an even (odd) number of fermions on the upper chain. 
The Hamiltonian allows for an exact derivation of the ground states if $u=1$ and $w=g\geq 0$, 
where the appearance of a robust ground state degeneracy characterized by the subchain
parity $\alpha$ as well as topological edge modes with Ising anyonic braiding statistics
has been demonstrated \cite{lang}.   
Notably, the ground state is the equal-weight superposition of all states with fixed 
total particle number $N$ and fermionic subchain parity $\alpha$.
The intuitive interpretation is the following:
The combination of intra-chain attraction with pair-hopping gives rise to $p$-wave pairing. 
Then, na\"ively, each wire acts as a mean-field superconductor for the other one
and allows for its description as Majorana chain \cite{kitaev_wires}. 
Remarkably, this picture remains valid despite the quasi long-range order in one-dimension.

At $u=0=g$, the model has previously been studied numerically with DMRG \cite{kraus}
where the appearance of edge modes has been observed away from half filling $\rho=N/2L\neq 1/2$. 
Therefore, Hamiltonian~(\ref{Hamiltonian}) can interpolate between the exactly solvable point of \cite{lang}
and the Hamiltonian describing a possible realization of a topological superconductor in cold atomic gases in \cite{kraus}.

%% ++++++++++++++++++++++++++++++++++++++++++++++++++++++++++++++++++++++++++++++++++++++++++++++++++++++++++++++++
\begin{figure}[t]
    \includegraphics[width=\columnwidth]{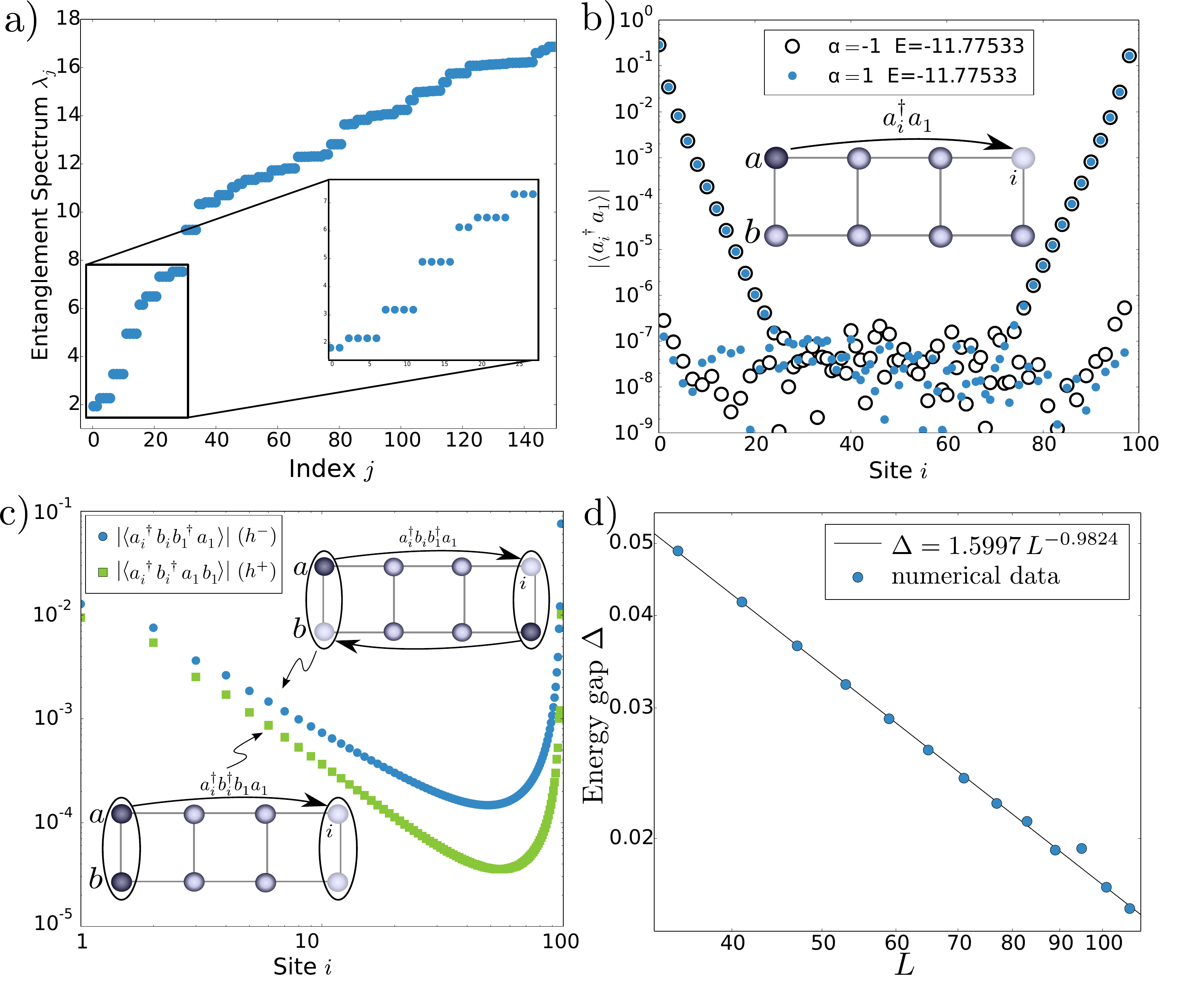}
    \caption{%
        \textit{Topological phase.}
        (a) Entanglement spectrum for a ladder of length $L=100$ with open boundaries and parameters $u=g=0.9$ and $w=1$ 
        at half filling with $\alpha=1$ for a bipartition at site
        $40$. The energy accuracy for the numerically determined
        approximate ground state is $\frac{\Delta E}{|E|} = 2.6\cdot 10^{-4}$.
        The twofold degeneracy is an indication of topological order. 
        (b) On-chain Green's function of this system featuring exponential decay and a revival at the other end of the chain, 
        indicating the existence of a localized edge state. For $\alpha=-1$,
        it is $\frac{\Delta E}{|E|} = 3.1\cdot 10^{-4}$. Note that the agreement in the
        ground state energy goes beyond accuracy because the energy error is
        expected to be the same for both states.
        (c) Inter-chain two-point correlation functions of the same state exhibiting algebraic behaviour and also a revival at the other end of the chain.
         (d) Scaling of the energy gap $\Delta$ in the topological phase for $u=g=0.9$ and $w=1$ 
        in the $\alpha=-1$ sector at half filling for open boundaries.
    }
    \label{fig:panelOne}
\end{figure}
%% ++++++++++++++++++++++++++++++++++++++++++++++++++++++++++++++++++++++++++++++++++++++++++++++++++++++++++++++++

%% ++++++++++++++++++++++++++++++++++++++++++++++++++++++++++++++++++++++++++++++++++++++++++++++++++++++++++++++++
\paragraph{Methods:} 
%% ++++++++++++++++++++++++++++++++++++++++++++++++++++++++++++++++++++++++++++++++++++++++++++++++++++++++++++++++

We analyze the ground state phase diagram with two different methods: 
In order to obtain a qualitative understanding of potentially competing phases, 
we apply the conventional bosonization methods \cite{mattis, vDelft} to obtain an effective low-energy field theory 
for the continuum limit of the model. 
On the other hand, we pin down the precise values for the phase transitions 
by performing a full scan of the phase diagram using DMRG.  

We start with the description of the bosonization procedure as reviewed in \cite{senechal}:  
The fermionic field operators $\psi_\sigma$ of chain $\sigma \in \{a,\,b\}$ in the continuum are
decomposed into left- and right-moving modes $\psi_{\sigma R/L}$ as
\begin{equation}
    \psi_\sigma(x) = \psi_{\sigma R}(x)\, e^{ik_\text{F}x} + \psi_{\sigma L}\, e^{-ik_\text{F}x}\,.
\end{equation}
Here, $k_\text{F}=\pi\rho$ is the Fermi-wavenumber defining the filling fraction $\rho$. 
The gist of bosonization is the expression of $\psi_{\sigma R/L}$ in terms of
the bosonic density field $\phi_\sigma$ and the phase field $\theta_\sigma$ via
\begin{equation}
    \psi_{\sigma R/L} = \frac{\eta_{\sigma R/L}}{\sqrt{2\pi}}\,
    :e^{-i\sqrt{\pi}\left(\theta_\sigma \pm \phi_\sigma \right)}:\,.
\end{equation}
Here, $\eta_{\sigma R/L}$ are Klein factors ensuring the anticommutation of
different fermion operators, and $:\bullet:$ denotes normal ordering.
It is convenient to introduce bosonic fields 
which are symmetric and antisymmetric with respect to chain exchange $a\leftrightarrow b$ \cite{cheng},
\begin{equation}
    \phi_\pm = \frac{1}{\sqrt{2}} \left( \phi_a \pm \phi_b \right)\,,
    \quad
    \theta_\pm = \frac{1}{\sqrt{2}} \left( \theta_a \pm \theta_b \right)\,.
\end{equation}
The term of the Hamiltonian responsible for the topological phase separates with respect to these sectors.
We then expand the full theory (\ref{Hamiltonian}) around the non-interacting point $u=g=w=0$
which is described by two decoupled Luttinger liquids.
The qualitative picture of the phase diagram follows from renormalization group arguments:
Including interactions leads to renormalized Luttinger parameters $K_\pm$ and, more importantly, 
provides potentially relevant terms which characterize transitions into new phases;
they will be discussed below.

For the quantitative mapping of the phase diagram, we study the model numerically 
using the well-established DMRG algorithm \cite{whiteA, whiteB} in its formulation as a variational matrix
product state ansatz \cite{schollwock}. 
Our implementation is tailored to Hamiltonian (\ref{Hamiltonian}) and 
allows for the efficient computation of ground- and low lying excited states,
exploiting the $U(1)$ symmetry (particle conservation) and the $\mathbb{Z}_2$ symmetry (subchain parity). 
To avoid local minima and enhance convergence, we employ subspace expansion \cite{hubig};
see the appendix for the utilized libraries.

%% ++++++++++++++++++++++++++++++++++++++++++++++++++++++++++++++++++++++++++++++++++++++++++++++++++++++++++++++++
\paragraph{Phase diagram:} 
%% ++++++++++++++++++++++++++++++++++++++++++++++++++++++++++++++++++++++++++++++++++++++++++++++++++++++++++++++++

We start analyzing the phase diagram at half filling, $N=L$, in the vicinity of
the exactly solvable point $u=g=w=1$. 
The \textbf{topological phase} (TP) is identified by three characteristic properties: 
(i) a robust ground state degeneracy with respect to the subchain parity $\alpha$, 
(ii) the existence of exponentially localized edge states manifest in a revival of the otherwise 
exponentially decaying on-chain Green's function $\langle a_i^\dag a_1\rangle$ at the end of the chain, 
and  (iii) a double degeneracy of the entanglement spectrum.  
In Fig.~\ref{fig:panelOne}~(a-b), we show DMRG results for these properties
at a generic point inside the topological phase.

\begin{figure}[t]
    \includegraphics[width=\columnwidth]{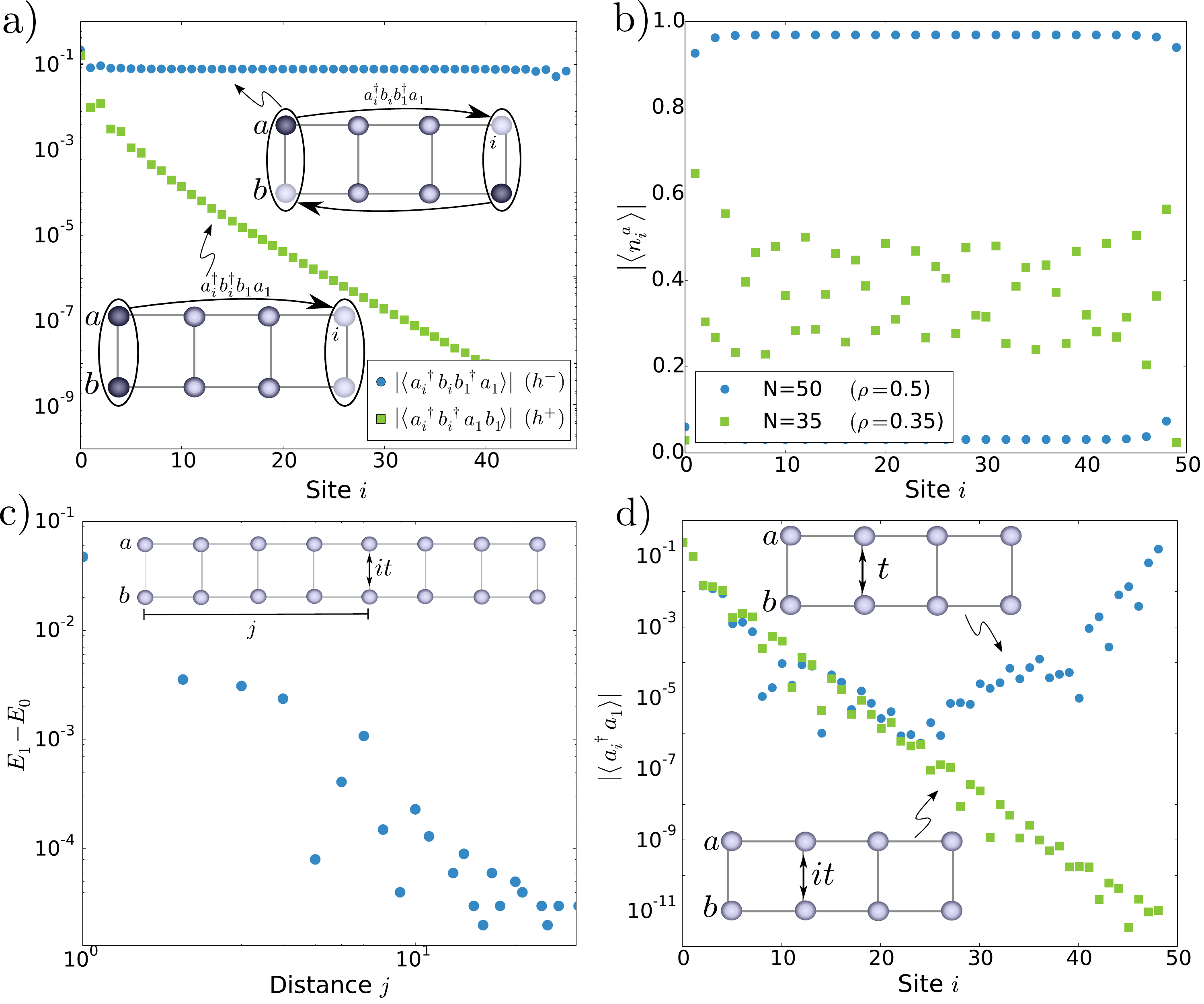}
    \caption{%
        (a) Inter-chain two-point functions $h^\pm$ for a chain of length $L=50$ with open boundaries for 
        $u=-0.5$, $g=-1$ and $w=1$ at half filling in the $\alpha=-1$ sector, 
        indicating the existence of a gap only in the symmetric
        sector. The energy accuracy is $\frac{\Delta E}{|E|} = 2.3\cdot 10^{-5}$.
      %  This behavior is a peculiarity at half filling and the exponential decay will fade for other filling fractions. 
        %
        (b) Local density for a chain of length $L=50$ with open boundaries for $u=g=-4$, $w=1$ in the $\alpha=-1$ sector indicating
        the charge density wave. Accuracy of energy is $\frac{\Delta E}{|E|} =
        6.8\cdot 10^{-8}\,(N=50)$ and $\frac{\Delta E}{|E|} =
        2.2\cdot 10^{-3}\,(N=35)$. Note that the accuracy for $N=50$ is much
        higher as the CDW+SDW state has an especially low entanglement entropy.
        (c) Dependence of the ground state energy splitting $E_1-E_0$ in the presence of a 
        local time-reversal breaking single-particle inter-chain hopping of strength $t=0.05$ 
        as a function of the distance $j$ from the edge. Results are  for an
        open chain of length $L=61$ with $N=43$ particles and $u=g=0.9$,
        $w=1$. Here, the energy error lies between
$\frac{\Delta E}{|E|} = 9.6\cdot 10^{-5}$ and $5.6\cdot 10^{-4}$.
        (d) On-chain Green's function for an open chain of length $L=50$ with $u=g=0.9$, $w=1$ at half filling 
        in the presence of \textit{global} single-particle hopping of strengh $t=0.05$ 
        with (circles, $\frac{\Delta E}{|E|} = 4.4\cdot 10^{-4}$) and without
        (squares, $\frac{\Delta E}{|E|} = 5.3\cdot 10^{-5}$) time-reversal symmetry.
    }
    \label{fig:panelTwo}
\end{figure}

These observations are in agreement with the qualitative results for the bosonized theory:  
Inter-chain pair hopping gives rise to the operator
\begin{equation}
    \psi^{\dagger}_{Ra}\psi^{\dagger}_{La}\psi^{}_{Lb}\psi^{}_{Rb} + \mathrm{h.c.}
    \;\propto\;
    \cos\left(\sqrt{8\pi/K_{-}}\;\theta_-\right)\,,
    \label{TopologicalRelevant}
\end{equation}
which becomes relevant for  $K_->1$. 
Then it is responsible for opening an energy gap in the asymmetric sector 
by pinning the relative phase $\theta_-$ to the values $\pm \sqrt{K_- \pi/8}$. 
As shown in Ref.~\cite{cheng}, the resulting effective theory describes a topological phase and
can be refermionized to the continuum version of the Majorana chain
\cite{kitaev_wires} at the Luther-Emery point $K_-=2$. Note that an
alternative approach for a topological phase has been studied in
Ref. \cite{keselman}, where the relevant term involves the density field
$\phi_-$ which corresponds to an interaction conserving the number of
particles on each wire.
The topological phase in the present manuscript directly connects to the exactly solvable, 
critical point for which the edge states have Ising anyonic braiding statistics \cite{lang}. 
This interpretation is confirmed by the numerical evaluation of the pair-correlation functions 
$h^-(x) = \langle \psi_a^{\dagger}(x) \psi_b (x)\psi_b^{\dagger}(0) \psi_a (0)\rangle$ 
and 
$h^+(x) = \langle \psi_a^{\dagger}(x) \psi_b^{\dagger}(x) \psi_b(0)\psi_a(0)\rangle$.
In the topological phase we find that both decay \textit{algebraically} 
and exhibit a revival at the edge, see Fig.~\ref{fig:panelOne}~(c). This observation
confirms the description of the topological phase  via attributing a mass to
the $\theta_-$ field.
Finally, the symmetric sector can be described by a Luttinger liquid with a linear low-energy spectrum. 
This behavior is confirmed by the DMRG analysis: 
The finite size gap in a system of length $L$ decays algebraically with $1/L$,
see Fig.~\ref{fig:panelOne}~(d). The variance of energy lies between $\frac{\Delta
E}{|E|}= 3.7\cdot 10^{-4}$ for the ground state calculations on short chains
and $1.3\cdot 10^{-3}$ for the
excited state calculations on long chains. Note that $\Delta E$ is a measure
of the maximal error of the gap. We can expect the results to be more accurate due to
error cancellation as DMRG is variational. Therefore the errors in the ground
state energy and the first excited state energies cancel to some extent when
computing the gap. The same argument applies to Fig.~\ref{fig:panelTwo}~(c).

This is in contrast to the exactly solvable point at  $u=g=w=1$, where the low-energy spectrum is quadratic and the gap 
closes as $1/L^2$ \cite{lang}. This exactly solvable point is therefore a critical point and characterizes the phase transition from
the topological phase towards phase separation, see below.

Decreasing the Luttinger parmeter $K_-<1$ eventually leads to a \textbf{gapless Luttinger liquid} (LL), 
where the operator (\ref{TopologicalRelevant}) becomes irrelevant. 
This phase is characterized by an algebraic decay of all correlation functions 
and a unique ground state, see Fig.~\ref{fig:PD}.

At half filling, two additional operators can become relevant and account for Umklapp processes. 
Both the inter- and intra-chain density-density interactions give rise to these processes
and two distinct and potentially relevant terms arise. 
The first one is the intra-chain Umklapp scattering 
\begin{equation}
    \left( \psi^{\dagger}_{R\sigma}\psi^{}_{L\sigma}\right)^2+\mathrm{h.c.}
    \;\propto\;
    \cos\left(\sqrt{8\pi K_{+}}\,\phi_+ \pm \sqrt{8\pi K_{-}}\,\phi_-\right)
\end{equation}
which becomes relevant for $(K_+ + K_-)<1$ and where the plus/minus sign corresponds to $\sigma=a/b$.
It leads to the formation of a \textbf{charge-density wave} (CDW+SDW) in both wires 
and provides a gap in the symmetric and antisymmetric sector, i.e., the whole theory is gapped.  
Indeed, the DMRG results show the appearance of this phase for strong intra-chain repulsion, see Fig.~\ref{fig:PD}, 
as expected from bosonization. 
Furthermore, we find both pair correlations 
$h^-(x)$
and 
$h^+(x)$
decaying exponentially, indicating that both $\phi_-$ and $\phi_+$ are well described as massive fields. 
The charge-density wave can also be directly seen in the local fermion density $\langle n_i^\sigma\rangle$, 
see Fig.~\ref{fig:panelTwo}~(b).

The second Umklapp process arises due to inter-chain density-density interactions 
\begin{equation}
    \left( \psi_{La}^{\dagger} \psi_{Ra}^{}\right)^2\left( \psi^{\dagger}_{Lb}\psi^{}_{Rb}\right)^2+\mathrm{h.c.} 
    \;\propto\;
    \cos\left(\sqrt{32\pi K_{+}}\,\phi_+\right)\,.
\end{equation}
It becomes relevant for $K_{+}<1/4$ and leads to a second type of \textbf{charge density wave} (CDW). 
Here, only the symmetric sector becomes gapped which manifests in only the correlation function 
$h^+(x)$
decaying exponentially whereas 
$h^-(x)$
shows algebraic behavior, see Fig.~\ref{fig:panelTwo}~(a). 
As expected, this phase appears for varying the strength of the inter-chain
interactions $g$. The transition from the topological to the LL and
CDW phases is described by a Kosterlitz-Thouless transition and hence of second
order; this is confirmed by the DMRG results.

Note that bosonization in principle would allow for an additional phase 
characterized by a spin density wave and driven by the operator
\begin{equation}
    \left( \psi_{La}^\dagger\psi^{}_{Ra}\right)^2\left( \psi_{Rb}^\dagger\psi^{}_{Lb}\right)^2 + \mathrm{h.c.}
    \;\propto\;
    \cos\left( \sqrt{32\pi K_-} \phi_- \right)\,.
\end{equation}
which originates from the inter-chain density-density repulsion
$n_i^an_i^bn_{i+1}^an_{i+1}^b$ contained in $B_i^2$. This operator is relevant for $K_-<\frac{1}{4}$. 
Recall that here the spin density wave is driven by a four-body interaction
whereas a general two-body inter-chain density-density interaction would become relevant for $K_-<1$.  
Within our DMRG analysis, we did not find such a spin density phase.

Finally, two types of \textbf{phase separation} are observed for strong attractive interactions: 
For strong intra-chain attraction ($u>0$), the phase separation occurs \textit{on} the chains (PS),
such that the fermions cluster on each chain while the particle number on both chains remains the same. 
The critical point $u=g=w=1$ is located on the phase boundary between the intra-chain phase separation and the topological phase.  
For a large, negative inter-chain density-density interaction $g$, on the other hand, one finds an inter-chain phase separation (F), 
as the interaction now favors all fermions clustering on one chain while the other chain is empty, 
see Fig.~\ref{fig:PD}. The transition from the topological phase to phase
separation is of first order, as supported by the numerical analysis.

%% ++++++++++++++++++++++++++++++++++++++++++++++++++++++++++++++++++++++++++++++++++++++++++++++++++++++++++++++++
%% ++++++++++++++++++++++++++++++++++++++++++++++++++++++++++++++++++++++++++++++++++++++++++++++++++++++++++++++++

%% ++++++++++++++++++++++++++++++++++++++++++++++++++++++++++++++++++++++++++++++++++++++++++++++++++++++++++++++++
\paragraph{Symmetry breaking and disorder:} 
%% ++++++++++++++++++++++++++++++++++++++++++++++++++++++++++++++++++++++++++++++++++++++++++++++++++++++++++++++++

Topological order manifests itself in a ground state degeneracy 
which is stable against disorder and symmetry-conserving perturbations. 
Using DMRG, we find that the ground state degeneracy and the degeneracy of the entanglement spectrum indeed are unaffected by disorder. 
Here, we implement disorder on the parameters $u,g$ and $w$ by making them site-dependent as 
$u_i = p_i u$ where $p_i \in [1-\delta, 1+\delta]$ are uniformly distributed random numbers and $\delta$ is the disorder strengh. 
The parameters $w$ and $g$ are treated analogously. 
For a moderate disorder of $\delta = 0.15$, 
the degeneracy of the entanglement spectrum remains unaffected in the topological phase, 
indicating its stability. 
The corresponding DMRG results are given in the appendix.

On the other hand, a natural perturbation breaking the subchain parity symmetry 
is single particle inter-chain hopping $a_i^\dag b_i+\mathrm{h.c.}$.
However, it was argued in Ref.~\cite{lang} that a residual time-reversal symmetry is sufficient to protect
the topological phase at the critical point $u=g=w=1$. 
This behavior is confirmed within our bosonization approach:
A generic, local single-particle hopping between the two chains takes the form
\begin{equation}
    P_\gamma(x) = 
    e^{i\gamma}\,\psi_a^\dagger (x) \psi_b^{} (x) 
    + e^{-i\gamma}\,\psi_b^\dagger (x) \psi_a^{} (x)
    \label{eq:hopping}
\end{equation}
with a phase factor $\gamma\in [0,2\pi)$.
In the topological phase, a time-reversal \textit{symmetric} perturbation ($\gamma=0$)
leads to exponentially decaying correlations $\left\langle P_\gamma(x)P_\gamma(0)\right\rangle$,
whereas a time-reversal symmetry \textit{breaking} perturbation ($\gamma\neq 0$) leads to an algebraic decay.
We can confirm this prediction within our DMRG analysis by studying the ground state properties
under perturbations of the form (\ref{eq:hopping}). 
Breaking time-reversal symmetry induces a ground state splitting which decays
with the distance of the perturbation from the edge, see
Fig.~\ref{fig:panelTwo}~(c).
In contrast, for time reversal symmetric hopping, even a homogenous perturbation leaves the topological phase intact,
with ground state degeneracy and revival of the Green's function at the edge, see Fig.~\ref{fig:panelTwo}~(d).
The numerical results for the topological phase impressively certify its
stability against perturbations violating subchain-parity conservation, as
expected from bosonization and predicted at the critical point \cite{lang}.
Finally, we would like to point out that for negative coupling $w$ the role
of these perturbations is reversed. This can be seen from the action of the
gauge transformation on the time-reversal symmetry operator.

%% ++++++++++++++++++++++++++++++++++++++++++++++++++++++++++++++++++++++++++++++++++++++++++++++++++++++++++++++++
\paragraph{Conclusion:} 
%% ++++++++++++++++++++++++++++++++++++++++++++++++++++++++++++++++++++++++++++++++++++++++++++++++++++++++++++++++

We investigated a model of interacting fermions in one dimension using
bosonization and DMRG, explicitly demonstrating the stability of the
topological properties of its exactly solvable, critical point
against certain perturbations. 
We numerically computed the phase diagram of the model,
featuring a variety of occuring phases that can be characterized by
bosonization. 
The stability of the topological properties was
demonstrated and we find them to be robust even in the presence of
perturbations violating subchain-parity conservation. Nevertheless, we find that
an additional symmetry (e.g., time-reversal) is required for an exponential splitting
of the ground state degeneracy, while a generic setup with only particle number conservation 
shows an algebraic splitting.

%% ++++++++++++++++++++++++++++++++++++++++++++++++++++++++++++++++++++++++++++++++++++++++++++++++++++++++++++++++
\begin{acknowledgments}
    We acknowledge support by the European Union under the ERC consolidator grant SIRPOL (grant N. 681208),
    and the Deutsche Forschungsgemeinschaft (DFG) within SFB/TRR 21.
    This research was supported in part by the National Science Foundation under Grant No. NSF PHY-1125915.
\end{acknowledgments}
%% ++++++++++++++++++++++++++++++++++++++++++++++++++++++++++++++++++++++++++++++++++++++++++++++++++++++++++++++++
\clearpage
\section{Appendix}
%% ++++++++++++++++++++++++++++++++++++++++++++++++++++++++++++++++++++++++++++++++++++++++++++++++++++++++++++++++
\paragraph{Bosonized Hamiltonian:}
%% ++++++++++++++++++++++++++++++++++++++++++++++++++++++++++++++++++++++++++++++++++++++++++++++++++++++++++++++++

To analyze the Hamiltonian qualitatively using bosonization, we separate the
original Hamiltonian (1) into
\begin{align}
    H = H_0 + H_1 + H_2 + H_3\,,
\end{align}
and treat the continuum limit of the lattice model to allow for the
field-theoretical description.
Here, $H_0$ is the kinetic part, describing free fermions with the well known
bosonized form~\cite{gogolin}
\begin{equation}
    H_0 = \frac{1}{2} \sum_{\eta=+,-} \int \mathrm{d}x \, (\partial_x \theta_{\eta})^2 + (\partial_x \phi_{\eta})^2\,,
\end{equation}
and $H_1$, $H_2$ and $H_3$ are the intra-chain density-density interaction,
the pair-hopping interaction and the inter-chain density-density interaction,
respectively. These interactions are now treated perturbatively using
bosonization and renormalization group arguments. The Fermi velocity is set to 1.

The bosonized form of the intra-chain density-density interaction is equally
well studied~\cite{senechal,vDelft} and has the bosonized form
\begin{align}
    H_{1} =& \frac{u}{4\pi} \sum_\sigma \int \mathrm{d}x\, \left( \frac{1}{2} +
    \sin^2(k_Fa) \right)(\partial_x \phi_\sigma)^2 \nonumber \\ &+  \left( \frac{1}{2} - \sin^2(k_Fa) \right)(\partial_x \theta_\sigma)^2 \,,
\end{align}
away from half filling. Here $k_F$ is the Fermi wavenumber encoding the
filling of the system. By rescaling of the fields, $H_1$ can be absorbed into
$H_0$~\cite{senechal} leading to
\begin{equation}
    \mathcal{H}_{0+1,c} = \sum_{\eta=\pm}\frac{\nu}{2} \left[ K_\eta (\partial_x \theta_{\eta})^2 + \frac{1}{K_\eta} (\partial_x \phi_{\eta})^2 \right]\,,
    \label{eq:luttinger_hamiltonian}
\end{equation}
with the new parameters 
\begin{subequations}
    \begin{align}
        K_\eta=& \sqrt{\frac{1+ \frac{u}{4\pi} (1 +2 \sin^2(k_Fa))}{1+ \frac{u}{4\pi} (1 - 2 \sin^2(k_Fa))}}\\
        \nu =& 2\sqrt{1+ \frac{2u}{4\pi} + \frac{u^2}{16\pi^2} (1-4\sin^4(k_Fa))}\,.
    \end{align}
\end{subequations}
The parameters $K_\pm$ are effectively treated by a rescaling of the fields
such that $H_{0}+H_{1}$ regains the form of $H_0$. That is,
\begin{subequations}
    \begin{align}
        \phi_{\pm} &\mapsto \frac{1}{\sqrt{K_\pm}} \phi_\pm\\
        \theta_\pm &\mapsto \sqrt{K_\pm}\theta_\pm\,.
    \end{align}
\end{subequations}
Note that although $K_+ = K_-$ while treating all interactions
perturbatively in lowest order, this is not strictly fulfilled; especially not
when going towards strong interactions, as can be seen from the DMRG results.

The pair hopping interaction term is known from studies of microscopic
models of topologically ordered systems~\cite{kraus, cheng} and has
(away from half filling) the bosonized form
\begin{align}
    \begin{split}
        \mathcal{H}_{2} =&\,q_1 \cos\left(\sqrt{8\pi/K_-}\theta_-\right)\\ 
        &+ q_2\cos\left(\sqrt{8\pi}\left(\sqrt{1/K_-} \theta_- + \sqrt{K_-}\phi_- \right) \right)\\
        &+ q_3\cos\left(\sqrt{8\pi}\left( \sqrt{1/K_-} \theta_- - \sqrt{K_-}\phi_- \right) \right)   \,,
    \end{split}
\end{align}
with $q_i\sim w$. When treating the interaction perturbatively and considering
the behaviour of the couplings $q_i$ under renormalization group, it turns out
that the second and third terms have scaling dimension $\Delta = 2\left( K_- +
\frac{1}{K_-}\right)$ and are therefore never relevant, whereas the first term
has scaling dimension $\Delta = \frac{2}{K_-}$ and becomes relevant for
$K_->1$. What remains is the famous sine-Gordon model~\cite{coleman}
which has been shown to refermionize to a system exhibiting toplogical order,
with the continuum version of the Kitaev chain being a special case at $K_-=2$~\cite{cheng}.

The inter-chain density-density interaction decomposes into intra-chain two-body
interactions which are of the same form as $H_1$ and can hence be absorbed
into $H_0$ by rescaling the fields, plus the three- and four-body
inter-chain interactions. 

We start from the bosonization of the full lowest
order continuum limit of the pair interaction $n_i^\sigma n_{i+1}^\sigma$ which is
\begin{equation}
    U_\sigma = V_\sigma + M_\sigma + F_\sigma\,,
\end{equation}
with the notation
\begin{subequations}
    \begin{align}
        \label{eq:V}
        V_\sigma =&  (\psi^{\dagger}_{R\sigma}\psi^{}_{R\sigma})^2 +
        (\psi^{\dagger}_{L\sigma}\psi^{}_{L\sigma})^2 \nonumber \\ 
        &+ 4\sin^2(k_Fa) \psi^{\dagger}_{R\sigma} \psi^{}_{R\sigma} \psi^{\dagger}_{L\sigma} \psi^{}_{L\sigma}\\
        M_\sigma =& \mathrm{e}^{2ik_Fx} \psi^{\dagger}_{L\sigma}\psi^{}_{R\sigma} (1+\mathrm{e}^{2ik_Fa}) +\mathrm{h.c.}\\
        F_\sigma =& \mathrm{e}^{4ik_Fx+2ik_Fa} \left( \psi_{L\sigma}^\dagger\psi_{R\sigma}\right)^2 + \mathrm{h.c.}\,.
    \end{align}
\end{subequations}
Away from half filling, the four-body interaction then reads in its bosonized form
\begin{align}
    H_3^{4-\text{body}} = & g \int\mathrm{d}x\,U_a(x)U_b(x) \nonumber\\
    = & \int \mathrm{d}x\,V_aV_b + M_aM_b + F_aF_b\,.
\end{align}
From equation (\ref{eq:V}) follows that $V_aV_b$ has scaling dimension
$\Delta = 4$ and is hence always irrelevant. When considering the three-body
term, we find that the relevant part is
\begin{equation}
    H_3^{3-\text{body}} = -g \int \mathrm{d}x\, M_aM_b\,.
\end{equation}
Thus, the contribution $M_aM_b$ precisely cancels and the inter-chain
density-density interaction reduces to the potentially relevant term
\begin{align}
    F_aF_b = &\left( \psi_{La}^\dagger\psi^{}_{Ra}\right)^2\left( \psi_{Rb}^\dagger\psi^{}_{Lb}\right)^2+\mathrm{h.c.}\nonumber\\
    \propto &\cos\left( \sqrt{32\pi K_-}\,\phi_- \right)\,.
\end{align}

In the special case of half filling, two additional relevant terms appear, as
certain oscillating terms do not average out anymore. 

These are the on-chain umklapp scattering 
\begin{align}
    &\left( \psi^{\dagger}_{R a/b}\psi^{}_{L a/b}\right)^2 + \mathrm{h.c.} \nonumber\\
\propto\;&\cos\left[\sqrt{8\pi}\left(\sqrt{K_+}\,\phi_+ \pm \sqrt{K_-}\,\phi_-\right)\right]\,,
\end{align} 
which comes from the intra-chain density-density interaction terms and has
scaling dimension $\Delta = 2(K_+ + K_-)$. In the domain of relevance, it
is driving the system towards charge-density wave order on both chains
simultaneously. 

The second term is the inter-chain umklapp scattering arising from
the four-body interaction which reads
\begin{align}
    &\left( \psi_{La}^{\dagger} \psi_{Ra}^{}\right)^2\left( \psi^{\dagger}_{Lb}\psi_{Rb}\right)^2 + \mathrm{h.c.}\nonumber\\
    \propto\;&\cos(\sqrt{32\pi K_+}\,\phi_+)\,.
\end{align}
It has scaling dimension $\Delta = 8K_+$ and can open a gap exclusively in the symmetric sector.

%% ++++++++++++++++++++++++++++++++++++++++++++++++++++++++++++++++++++++++++++++++++++++++++++++++++++++++++++++++
\paragraph{Evaluation of the phase diagram:}
\label{app:color}
%% ++++++++++++++++++++++++++++++++++++++++++++++++++++++++++++++++++++++++++++++++++++++++++++++++++++++++++++++++

The color code of Fig.~1 in the main text is constructed from the DMRG ground
state calculations in both subchain parity sectors using the gap between the sectors, $|E_0^{\alpha = 1} - E_0^{\alpha = -1}|$, 
the ratio of minimum to maximum on-chain density $\Braket{n_i^a}$
\begin{equation}
    \Delta n = \frac{\mathrm{min}\left\lbrace\Braket{n_i^a}\right\rbrace}{\mathrm{max}\left\lbrace\Braket{n_i^a}\right\rbrace}\,,
\end{equation}
the revival of the Green's function
\begin{equation}
    v = \frac{\Braket{c_L^\dagger c_1^{}}}{\Braket{c_1^\dagger   c_1^{}}}\,,
\end{equation} 
the minimum of that correlation function's envelope $m$, 
the mean fluctuation of $\Braket{n_i^a}$,
\begin{equation}
    f = \Braket{|\Braket{n_i^a} - \Braket{\Braket{n_i^a}}_i|}_i\,,
\end{equation}
where $\langle \rangle_i$ denotes the spatial average, the mean minimal
density
\begin{equation}
    a = \Braket{\mathrm{min}\left\lbrace n_i^a, n_i^b \right\rbrace}_i\,,
\end{equation}
and the standard deviation of energy $\Delta E = \sqrt{\Braket{H^2} - \Braket{H}^2}$. 
The two-point correlation functions $h^\pm$ are not directly
employed, but checked explicitly at exemplary points within the phases. 
The color $(r,g,b)$ in RGB code is now attributed as follows: 
\begin{enumerate}
    \item If $f>\frac{1}{4}$ and $\Delta n < \frac{1}{50}$, we face a large variance in density
        that extends over the entire system. As all observed cases of charge-density
        waves have significantly larger $\Delta n$, 
        we can safely attribute on-chain phase separation and set $(r,g,b) = (0,1,0)$. 
    \item If $f<\frac{3}{10}$ and $a<\frac{1}{5}$, there is little density fluctuation on the
        chains but a large difference between them, which qualifies for
        inter-chain phase separation and we set $(r,g,b) = (1,0,1)$.
    \item If neither on- nor inter-chain phase separation are found, we set the
        color as follows:
        \begin{align}
            r &= \mathrm{e}^{-|E_0^{\alpha = 1} - E_0^{\alpha = -1}|} \nonumber \\
            g &= \begin{cases} 0 & \mathrm{if}\quad m > \frac{1}{4} \\ 7v & \mathrm{if}\quad m<\frac{1}{4},\;
                v<\frac{1}{10} \\ \frac{7}{10}+\frac{3}{10}\sin \left( \frac{\pi (v-\frac{1}{10})}{\frac{18}{10}}\right)&
                \mathrm{else}  \end{cases} \nonumber \\
            b &= \begin{cases} 1 & \mathrm{if}\quad f>\frac{3}{5} \\ \frac{5}{3}f & \mathrm{else} \end{cases}
        \end{align}
\end{enumerate}
The threshold values are empirically determined by explicitly considering the
decisive correlations and densities in the respective domains in parameter
space.

The saturation is then used to represent the accuracy in form of the standard deviation of energy
$\Delta E$: It is set to $\mathrm{e}^{-30\Delta E}$, that
is, if the result is an exact eigenstate, it is $\Delta E = 0$ and therefore
the saturation is 1, else it decays exponentially with the standard deviation. 

Using this coloring scheme, we can visualize the different phases, as the
phase separation is identified directly, the CDW+SDW phase features large
on-chain density fluctuations but neither revival nor a degeneracy with
respect to $\alpha$ and thus appears blue, the CDW phase does have a
homogeneous density and no revival, but the groundstate is degenerate with
respect to $\alpha$, it hence appears red. The topological phase also has a
homogeneous density, but features revival and ground state degeneracy and
therefore is colored yellow. The LL phase does not have any of the probed
characteristics and hence appears grey.

%% ++++++++++++++++++++++++++++++++++++++++++++++++++++++++++++++++++++++++++++++++++++++++++++++++++++++++++++++++
\paragraph{Additional DMRG results:}
\label{app:data}
%% ++++++++++++++++++++++++++++++++++++++++++++++++++++++++++++++++++++++++++++++++++++++++++++++++++++++++++++++++

%% ++++++++++++++++++++++++++++++++++++++++++++++++++++++++++++++++++++++++++++++++++++++++++++++++++++++++++++++++
\begin{figure}[t]
    \includegraphics[width=\columnwidth]{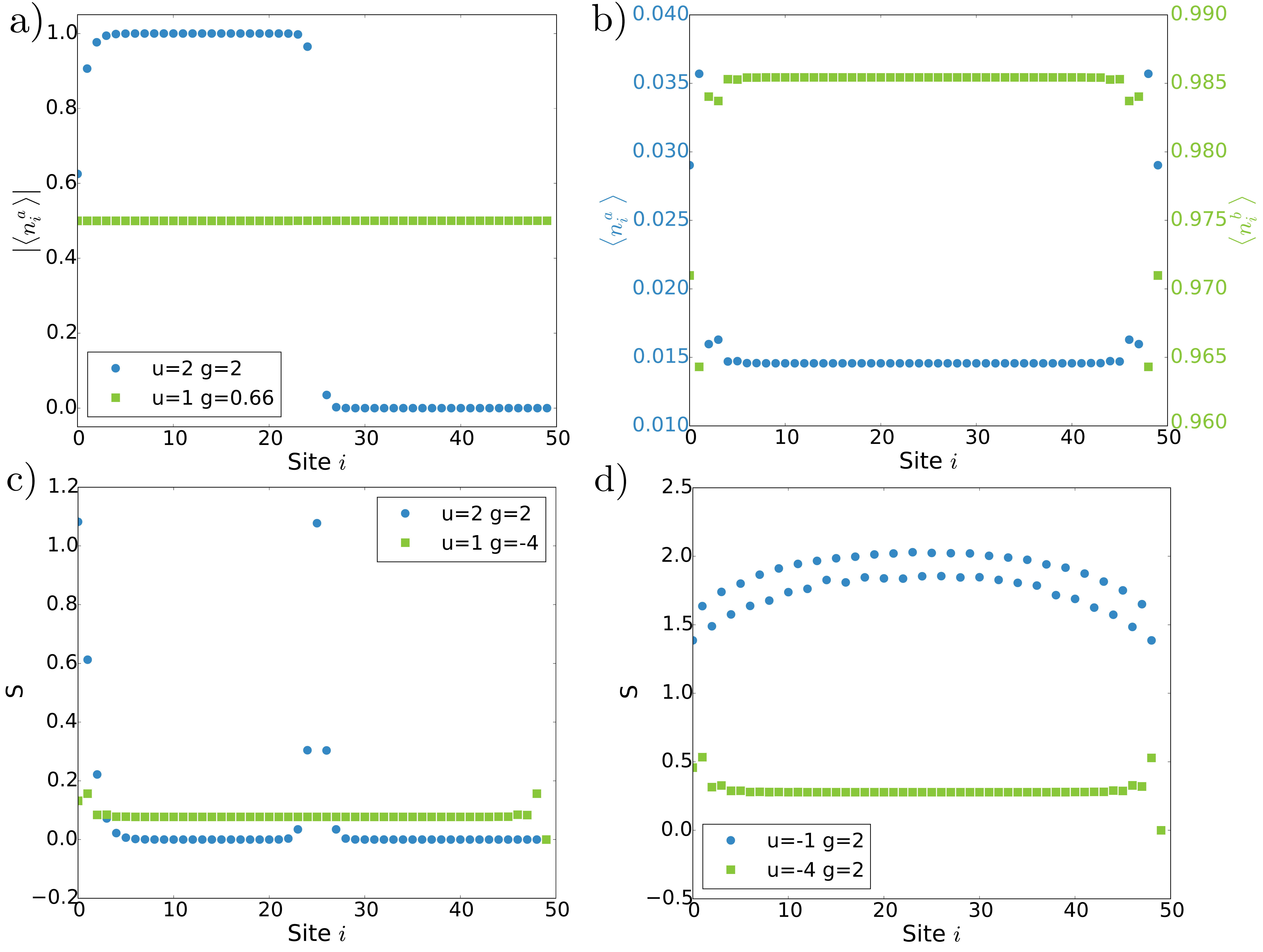}
    \caption{(a) Local fermion density on one chain for the topological
        phase at $w=1,\,u=1,\,g=0.66$ $(\frac{\Delta E}{|E|} = 0.0124)$ and
        the phase separation at $w=1,\,u=g=2$ $(\frac{\Delta E}{|E|} = 2.676\,10^{-8})$ at half filling
        for two chains of length $50$. With the fermion
        numbers being close to $1$ and $0$ in the phase separation, these values can
        be interpreted as eigenvalues. (b) Local fermion densities on both
        chain $a$ (blue) and $b$ (green) in the inter-chain phase separation at
        $w=1,\,u=1,\,g=-4$ at half filling for two chains of length $50$ $(\frac{\Delta E}{|E|} = 1.375\,10^{-5})$. 
        Again, the fermion numbers being close to $1$ and $0$ gives a
        low entanglement phase. (c) Entanglement entropy in the phase
        separation for on-chain (blue) and inter-chain (green) phase separation as a
        function of the bipartition site. The
        peaks in the entanglement entropy of the on-chain phase separated state stem
        from the transition between the entirely filled to the empty
        region. (d) Entanglement entropy in the full charge-density wave
        (green) $(\frac{\Delta E}{|E|} = 4.625\,10^{-8})$ and in the Luttinger liquid (blue) $(\frac{\Delta E}{|E|} =
        4.969\,10^{-4})$ as a function of the bipartition site.}
    \label{fig:appendix1}
\end{figure}
%% ++++++++++++++++++++++++++++++++++++++++++++++++++++++++++++++++++++++++++++++++++++++++++++++++++++++++++++++++

%% ++++++++++++++++++++++++++++++++++++++++++++++++++++++++++++++++++++++++++++++++++++++++++++++++++++++++++++++++
\begin{figure}[t]
    \includegraphics[width=\columnwidth]{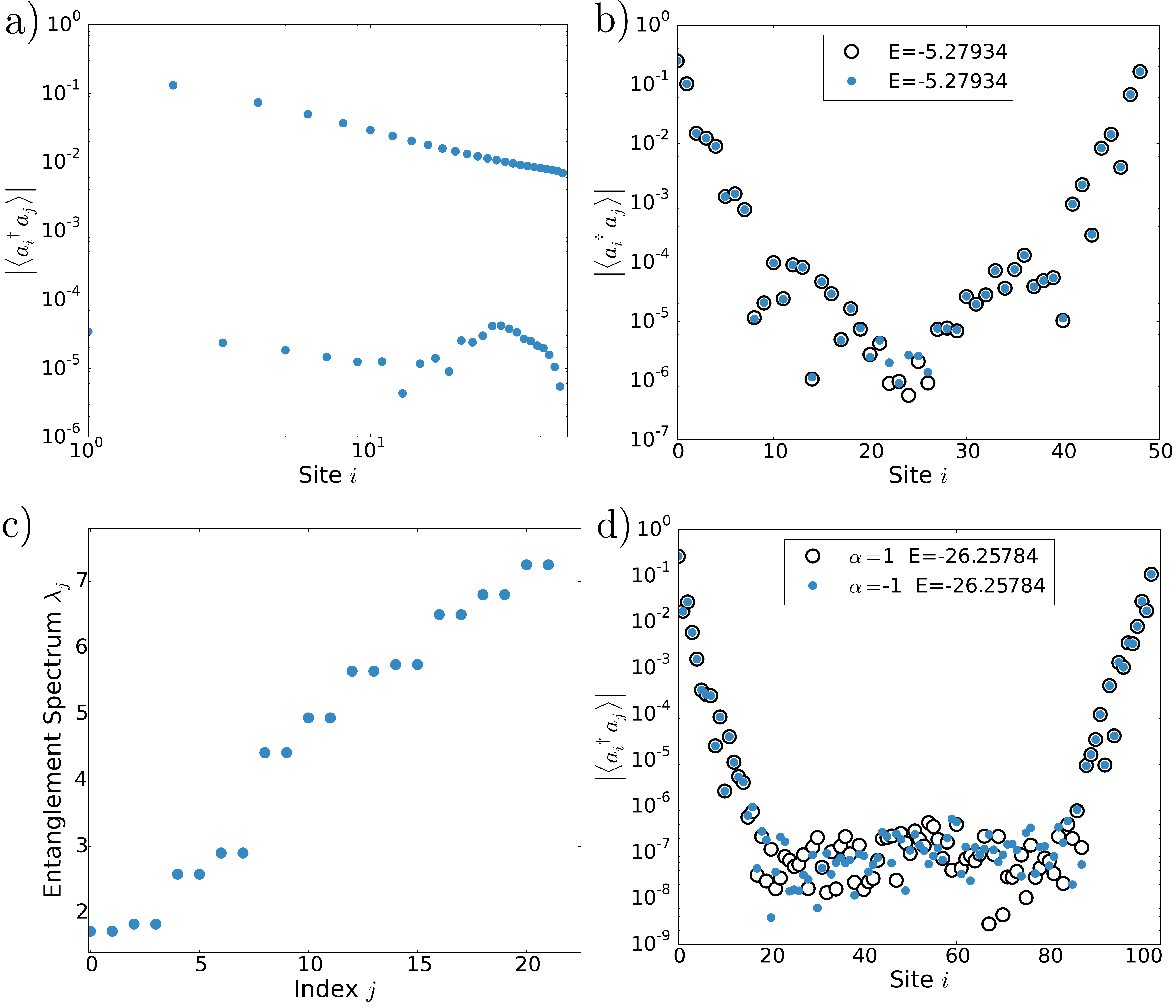}
    \caption{(a) On-chain Green's function in the Luttinger
        liquid phase at $w=1,\,u=-1,\,g=2$ at half filling for two chains of
        length $50$ $(\frac{\Delta E}{|E|} =
        4.969\,10^{-4})$. 
        The oscillatory behaviour of the correlation
        function leads to half of the datapoints being located at low values, but
        the algebraic decay of the envelope is clearly visible which connects to
        the phase being gapless in both sectors. (b) On-chain Green's function
        in the topological phase at $w=1,\,u=g=0.9$ for two chains of
        length $50$ with a fermion number of $35$ in the presence of a global
        single-particle hopping of strengh $t=0.05$ $(\frac{\Delta E}{|E|} = 4.43\,10^{-4})$. (c) Entanglement spectrum
        for the topological phase at $w=1.1,\,u=g=0.8$ $(\frac{\Delta E}{|E|} =
        3.32\,10^{-4})$ for a bipartition at site $i=64$ with
        the total chain length $104$ and the particle number $93$. Here, a
        disorder of $\delta = 15\%$ in all parameters is applied.
        %, that is, all parameters vary spatially with a random equally distributed noise of $\delta$. 
        The double degeneracy is clearly visible, indicating stability
        against spatial disorder. (d) Green's function and ground state degeneracy for
    the same system, showing the robustness of the edge state against disorder.}
    \label{fig:appendix2}
\end{figure}
%% ++++++++++++++++++++++++++++++++++++++++++++++++++++++++++++++++++++++++++++++++++++++++++++++++++++++++++++++++

The data presented here illustrates some of the properties and
characteristics of the phases. 
The phase separation (both on the chains and between the chains) can directly be seen from the local fermion density
$\Braket{n_i^{a/b}}$, as shown in Fig.~\ref{fig:appendix1} (a) and
(b). The entanglement entropy in these states is extremely low, as
shown in Fig.~\ref{fig:appendix1} (c), such that they can be pictured
as being close to product states and hence eigenstates of the local particle
number operators. Note that such product states cannot be eigenstates of the
model Hamiltonian due to the kinetic term, but the energy variance is
sufficiently low here [with $(\Delta E)^2 = 6.28\,10^{-9}$ for the on-chain phase separation
    and $(\Delta E)^2 = 7.13\,10^{-10}$ for the inter-chain phase separation]
so that these states are close to the true groundstate.

The area law for entanglement entropy~\cite{eisert} in one-dimensional systems
demands a constant entanglement entropy for gapped, one-dimensional
Hamiltonians. As the bosonization predicts a full gap in the CDW+SDW phase, we
therefore expect a constant entanglement entropy in this case, which is found in
DMRG calculations as depicted in Fig.~\ref{fig:appendix1} (d). For
comparison, we also show the entanglement entropy of the gapless Luttinger
liquid phase that clearly varies with the subsystem size.

Also, the Luttinger liquid phase is characterized by a non-degenerate groundstate,
as shown in the phase diagram in Fig.~1 of the main text, and algebraically
decaying correlations as shown in Fig.~\ref{fig:appendix2} (a) --- which
corresponds to the system being gapless in both sectors. 

Adding a weak local single-particle inter-chain hopping does not break the
topologically ordered phase. This is not limited to the
perturbation being local: A global single-particle hopping has the same effect,
as shown in Fig.~\ref{fig:appendix2} (b), i.e., both the ground state
degeneracy and the exponentially localized edge states remain stable in the
presence of global weak single-particle inter-chain hopping. This is not
the case if the single-particle hopping breaks time-reversal symmetry; then
even a local perturbation can destroy the edge states.

The stability of the topological phase against disorder manifests directly in
the stability of the degeneracy of the ground states and the entanglement spectrum as
well as the exponential decay of the Green's function with a revival at the
other end of the chain. This can be seen exemplarily in Fig.~\ref{fig:appendix2}
(c) and (d), showing the entanglement spectrum and the Green's
function which features the characteristic revival even in the presence of
disorder.

%% ++++++++++++++++++++++++++++++++++++++++++++++++++++++++++++++++++++++++++++++++++++++++++++++++++++++++++++++++
\paragraph{DMRG Implementation:}
\label{app:dmrg}
%% ++++++++++++++++++++++++++++++++++++++++++++++++++++++++++++++++++++++++++++++++++++++++++++++++++++++++++++++++

We implemented the DMRG algorithm using \texttt{C++11} and utilizing the
\texttt{BLAS}~\cite{blas}, \texttt{LAPACK}~\cite{lapackug} and \texttt{Arpack}~\cite{arpack} libraries. 
We therefore make use of the \texttt{BLAS} and \texttt{LAPACK} routines and their respective \texttt{C} interface from
\texttt{Intel\textsuperscript{\textregistered} MKL} and the \texttt{Arpack++ C++} wrapper for \texttt{Arpack}~\cite{arpack++}. 
The code is parallelized for shared memory architectures using \texttt{openMP}.

%% ++++++++++++++++++++++++++++++++++++++++++++++++++++++++++++++++++++++++++++++++++++++++++++++++++++++++++++++++

%% ++++++++++++++++++++++++++++++++++++++++++++++++++++++++++++++++++++++++++++++++++++++++++++++++++++++++++++++++

\bibliographystyle{ieeetr}
%\input{preprint.bbl}

%\clearpage
%\input{appendix.tex}

\end{document}